\documentclass[9pt,twocolumn,twoside]{osajnl}

\journal{pr} % Choose journal (ao,jocn,josaa,josab,ol,optica,pr)

\setboolean{shortarticle}{false}

\usepackage{subcaption}
\usepackage{graphicx}
\usepackage{lineno}
\usepackage{lipsum} %%% 
\usepackage{physics}
\usepackage{amsmath}
\usepackage{amssymb}
\usepackage{gensymb}
\title{Finite-size security of continuous-variable quantum key distribution with imperfect heterodyne measurement}

\author[1,**,$\dagger$]{Adnan A.E. Hajomer}
\author[2,**]{Akash nag Oruganti}
\author[1, $\ddagger$]{Ivan Derkach}
\author[1]{Ulrik L Andersen}
\author[2]{Vladyslav C Usenko}
\author[1,$\ddagger$]{Tobias Gehring}

\affil[1]{Center for Macroscopic Quantum States (bigQ), Department of Physics, Technical University of Denmark, 2800 Kongens Lyngby, Denmark}

\affil[2]{Department of Optics, Faculty of Science, Palacky University, 17. listopadu 50, 772 07 Olomouc, Czech Republic}

\affil[ ]{Corresponding author: aaeha@dtu.dk, $^\dagger$ tobias.gehring@fysik.dtu.dk,$^{\ddagger}$ ivan.derkach@upol.cz}

\affil[**]{\textbf{These authors contributed equally as first authors}}

\begin{abstract}
Continuous-variable quantum key distribution (CVQKD) using coherent states and heterodyne detection enables secure quantum communication based on technology that has large similarities to coherent optical telecommunication practical implementations of coherent receivers used in both technologies encounter device imperfections, which for CVQKD are often not addressed in security proofs. Here, we present a theoretical framework that rigorously accounts for imperfect heterodyne measurements arising from phase imbalances in the coherent (heterodyne) receiver. Focusing on collective attacks, we establish a finite-size security proof that reveals how measurement imperfections limit the distance over which a positive key rate is achievable. To mitigate these effects, we propose a local transformation during classical post-processing.
We validate our approach experimentally on a CVQKD system with an imperfect coherent receiver, underscoring its potential for scalable, cost-effective CVQKD with photonic integrated receivers in which phase-imbalances naturally appear through manufacturing tolerances.   
\end{abstract}

\setboolean{displaycopyright}{true}

\begin{document}

\maketitle

\section{Introduction}  \label{chapter:intro}

Quantum key distribution (QKD) is one of the most advanced technologies emerging from quantum information theory, with the potential for widespread commercialization~\cite{pirandola2020advances}. However, for large-scale deployment, next-generation QKD systems must employ low-cost, integrated photonic devices capable of high-rate performance and seamless integration with existing and emerging classical network infrastructure~\cite{sibson2017integrated,zhang2019integrated,zhu2022experimental,hajomer2024continuous}. 

Continuous variable (CV) quantum key distribution (QKD), where key information is encoded into the two orthogonal quadratures—phase and amplitude—of a coherent state, is especially suitable for photonic integration~\cite{hajomer2024continuous,zhangIntegratedSiliconPhotonic2019b}. In the CVQKD protocol, the sender (Alice) uses a quadrature modulator to prepare coherent states and sends them through an insecure quantum channel that is assumed to be under the full control of an eavesdropper (Eve). The receiver (Bob) decodes this quantum information through coherent detection methods, such as homodyne or heterodyne detection, facilitated by a local oscillator (LO) ~\cite{kikuchi2015fundamentals, weedbrook2012gaussian}. The CVQKD protocol based on heterodyne detection, which is the joint measurement of the two quadratures (referred to as the no-switching protocol~\cite{weedbrook2004quantum}), offers the most advanced security proof and an implementation advantage through the heavy use of digital-signal-processing~\cite{leverrier2015composable, pirandola2024improved, jain2022practical, Kanitschar2023, hajomer2024experimental}.

Heterodyne detection can be achieved via a phase-diverse homodyne receiver or an RF heterodyne technique~\cite{kikuchi2015fundamentals,kleis2017continuous,jain2022practical,qi2015generating,ren2021demonstration,laudenbach2019pilot}. For broadband CVQKD implementations, phase-diverse homodyne receivers are particularly practical, as they require quantum-shot-noise-limited detectors with only half the bandwidth of RF heterodyne techniques~\cite{hajomer2023continuous,ren2021demonstration,laudenbach2019pilot,wang2022sub}. Additionally, phase-diverse homodyne receivers can be integrated using silicon photonics~\cite{hajomer2023continuous}, allowing compatibility with CMOS technology~\cite{doerr2018silicon}.

However, integrated CVQKD phase-diverse homodyne receivers must meet specific requirements distinct from those of classical optical telecommunications receivers. For instance, CVQKD applications demand orthogonal measurements with minimal phase error (i.e., phase-imbalance-free) and a balanced splitting ratio (amplitude-imbalance-free) for the quantum signal. Despite these demands, integrated phase-diverse receivers tend to exhibit phase imbalances, typically around $3^{\circ}$ to $10^{\circ}$ for $90^{\circ}$ optical hybrids based on multi-mode interference (MMI) couplers, and amplitude imbalances due to manufacturing tolerances~\cite{van2014tolerant}. Therefore, to facilitate the development of low-cost, integrated CVQKD receivers, it is essential to explore theoretical approaches or compensation methods that can account for or mitigate these imperfections.

In this work, we present a theory that quantifies the secret key rate achievable in Gaussian modulated coherent state CVQKD using imperfect heterodyne measurements. Unlike previous studies~\cite{lupo2022quantum,pereira2021impact}, our theory incorporates phase imbalance within the CVQKD phase-diverse homodyne receiver and focuses on collective attacks in the finite-size regime. We demonstrate that phase imbalance can significantly reduce the mutual information between Alice and Bob, while the Holevo bound increases due to misidentifying part of the signal from the conjugate quadrature as noise. This imperfection consequently limits the distance over which a secret key rate can be achieved. Furthermore, we propose a compensation method that employs a local transformation on the generated quantum symbols. The efficacy of this compensation method is validated through a proof-of-principle experiment using a 90° optical hybrid with a 10° phase imbalance, which is typical for the fabrication tolerances of photonic integrated circuits~\cite{van2014tolerant}. Our positive results demonstrate that the application of this local transformation can enhance the secret key rate for imperfect receivers. This advancement may relax the stringent requirements for integrated CVQKD receivers, facilitating large-scale deployment of CVQKD.

%%%%%%%%%%%%%%%%%%%%%%%%%
    
    \section{Theoretical background} \label{chapter:theory}
A typical prepare-and-measure (PM) coherent-state Gaussian CVQKD protocol involves Alice preparing an initial coherent state with canonical quadratures $x(p)_s$, and modulating the state (using an amplitude and phase, or an IQ modulator \cite{jain2021modulation}) with the modulation strength $\alpha$ according to random variables $x(p)_a$ drawn from two independent zero-mean Gaussian distributions with variance $V^{x(p)}_A$. The resulting quantum state, characterized in respective quadratures as $x(p)_A=x(p)_s+\alpha x(p)_a$ (with variances $1+\alpha^2V^{x(p)}_A$), is sent through an untrusted quantum channel, where it is exposed to losses $\eta$ and noise $x(p)_\varepsilon$, the latter having respective variances $\varepsilon_{x(p)}$ \cite{grosshans2002continuous}, to Bob. Bob then proceeds to the measurement of both quadratures of the received signal using a balanced heterodyne detector. 
Disregarding limited detector efficiencies and electronic noise (for now), such measurement will yield respectively $x_B=\sqrt{\eta/2}(x_s+\alpha x_t+x_{\varepsilon})+\sqrt{1-\eta/2}x_0$ and $p_B=\sqrt{\eta/2}(p_s+\alpha p_t+x_{\varepsilon})-\sqrt{1-\eta/2}p_0$, where $x(p)_0$ corresponds to quadratures of the added vacuum noise. The overall variance of the received state is $V^{x(p)}_B$.
The covariance matrix of the random vector $\{x_a,p_a,x_b,p_b\}$ is given by 
   \begin{equation}\label{eq:cov-matrix}
        \Gamma=
        \begin{bmatrix}
        \gamma_A & \gamma_C \\
        \gamma_C^T & \gamma_B \\
        \end{bmatrix}=\left[\begin{array}{cc|cc}
        V^x_A & \varsigma_{A^x,A^p}& \sigma^x_{A,B} & \varsigma_{A^x,B^p} \\
        \varsigma_{A^x,A^p} & V^p_A & \varsigma_{A^p,B^x} & \sigma^p_{A,B}\\ \hline
        \sigma^x_{A,B}& \varsigma_{A^p,B^x} & V^x_B & \varsigma_{B^x,B^p}\\
        \varsigma_{A^x,B^p}& \sigma^p_{A,B} & \varsigma_{B^x,B^p} &V^p_B 
    \end{array}\right].    \end{equation}
By definition $\varsigma_{A^x,A^p}=0$, while other anti-diagonal elements $\varsigma_{A^x,B^p}$, $\varsigma_{A^p,B^x}$ and $\varsigma_{B^x,B^p}$ are expected to be negligible (comparing to co-variances $\sigma^{x(p)}_{A,B}$) and commonly disregarded~\cite{jouguet2012analysis}. The matrix (\ref{eq:cov-matrix}) is sufficient to evaluate the mutual information between the trusted parties. However, it does not directly provide the eavesdropper's accessible information. To ascertain this, one must understand the underlying cause of the imperfections, that is, model a system that could produce them (see Sec.\ref{sec:modelling}). Once the imperfections are modeled, the modulated coherent state is replaced by a two-mode squeezed vacuum \cite{laudenbach2018continuous} to obtain the Holevo bound, which serves as an upper limit on the eavesdropper's information. Consequently, it allows us to lower bound the secure key rate \(K\), defined as the information advantage of the trusted parties \cite{devetak2005distillation}.
 In the asymptotic regime, the rate of the secret key, secure against collective attacks, is 
    \begin{equation}
        K\left(\eta,\varepsilon \right)\ge\text{max}\left[0,\beta I(A\!:\!B)- \chi(E)\right],
            \label{eq:skr-asymptotic}
    \end{equation}
where reconciliation efficiency $\beta\in(0,1)$ indicates the inability to extract $I(A\!:\!B)$ in both quadratures completely. In the following section we inspect the evaluation of the mutual information  (MI) under general considerations, as well as approaches and obstacles in its recovery. \\

%%%%
    \subsection{Mutual information}\label{sec:mutual-info}
Upon correct measurement and recovery of encoded symbols the information carried in each quadrature will be independent implying that the block matrices $\gamma_A,\,\gamma_B,\,\gamma_C $ will remain diagonal. In this case the MI can be assessed using the signal-to-noise ratio (SNR) estimated by considering the quadratures independently, $I(A\!:\!B)=1/2\log_2(1+ \text{SNR})$ \cite{guo2005mutual, laudenbach2018continuous}, or equivalently directly from the covariance matrix (\ref{eq:cov-matrix}):
    \begin{equation}\label{eq:mi-basic}
        I(A\!:\!B)=\frac{1}{2}\left( \log_2 \frac{V^x_A}{V^x_{A|B}}+\log_2 \frac{V^p_A}{V^p_{A|B}}\right),
    \end{equation}
where $V^{x(p)}_{B|A}$ are variances of Bob conditioned on Alice's measurement, which in the simplest case are given as
\begin{equation*}
    V^{x(p)}_{A|B}=V^{x(p)}_A-\frac{\left(\sigma^{x(p)}_{A,B}\right)^2}{V^{x(p)}_B}.
\end{equation*}
However, due to various practical imperfections, the block matrices $\gamma_{A,B,C}$ are not diagonal. %which is predominantly caused by limited practical phase matching, which reduces correlations $\sigma$ \cite{hajomer2024long}. 
The security analysis can be simplified by \textit{symmetrization} of the matrix $\Gamma$ \cite{leverrier2015composable}, i.e.\ by ignoring all $\varsigma$ elements and retaining reduced correlations $\sigma$ in Eq.(\ref{eq:cov-matrix}). This does not compromise security as reducing correlations $\sigma$ decreases $I(A\!:\!B)$ and increases the eavesdropper's information $\chi(E)$. We refer to the MI obtained from a symmetrized covariance matrix $I(A\!:\!B)_{\varsigma\to0}$ as \textit{ignorant} mutual information. %Consequently, the cost of such simplification is a significantly diminished key rate (\ref{eq:skr-asymptotic}). 
Such simplification streamlines the security analysis, but significantly diminishes the performance of the system~\cite{leverrier2015composable}. On the other hand, a more general approach to evaluation of the MI allows to achieve better performance (see Sec:\ref{sec:mutual} in the Supplementary Materials for details): 
    \begin{equation}\label{eq:mi-full}
        I_T(A\!:\!B)=\frac{1}{2}\log_2\frac{|\gamma_A|}{|\gamma_{A|B}|}\ ,
    \end{equation}
where $|\cdot|$ is the determinant of the respective matrix, and $\gamma_{A|B}$ is Alice's matrix conditioned by the measurement of Bob \cite{weedbrook2012gaussian}. We refer to the MI in Eq.(\ref{eq:mi-full}) as \textit{true} MI, as it does not treat independent modulation in the complementary quadrature as noise and is the maximum achievable MI. The true MI can be expressed in terms of SNR (in this context, the SNR is defined by recognizing the signal from both quadratures simultaneously rather than treating them as independent entities), but requires also the MI between Bob's quadratures and the MI between quadratures of Bob's conditional state (see Supplementary Materials for derivation)
    \begin{equation}\label{eq:mut-snr}
        I_T(A\!:\!B)=\log_2(1+SNR)-I(B^x:B^p)+I(B^x|A:B^p|A)\ .
    \end{equation}
Consequently, we have two distinct expressions for the MI -- the true MI $I_T(A\!:\!B)$ from Eq.(\ref{eq:mi-full}), and the ignorant $I(A\!:\!B)_{\varsigma\to 0}$, where clearly, the latter is more conservative as $I_T(A\!:\!B)\geq I(A\!:\!B)_{\varsigma\to 0}$ and equality holds in absence of practical imperfections. %phase mismatch.

    \subsection{Transformations to aid information reconciliation}\label{sec:transformation}

\begin{figure*}
        \centering
    \includegraphics[width=.7\linewidth]{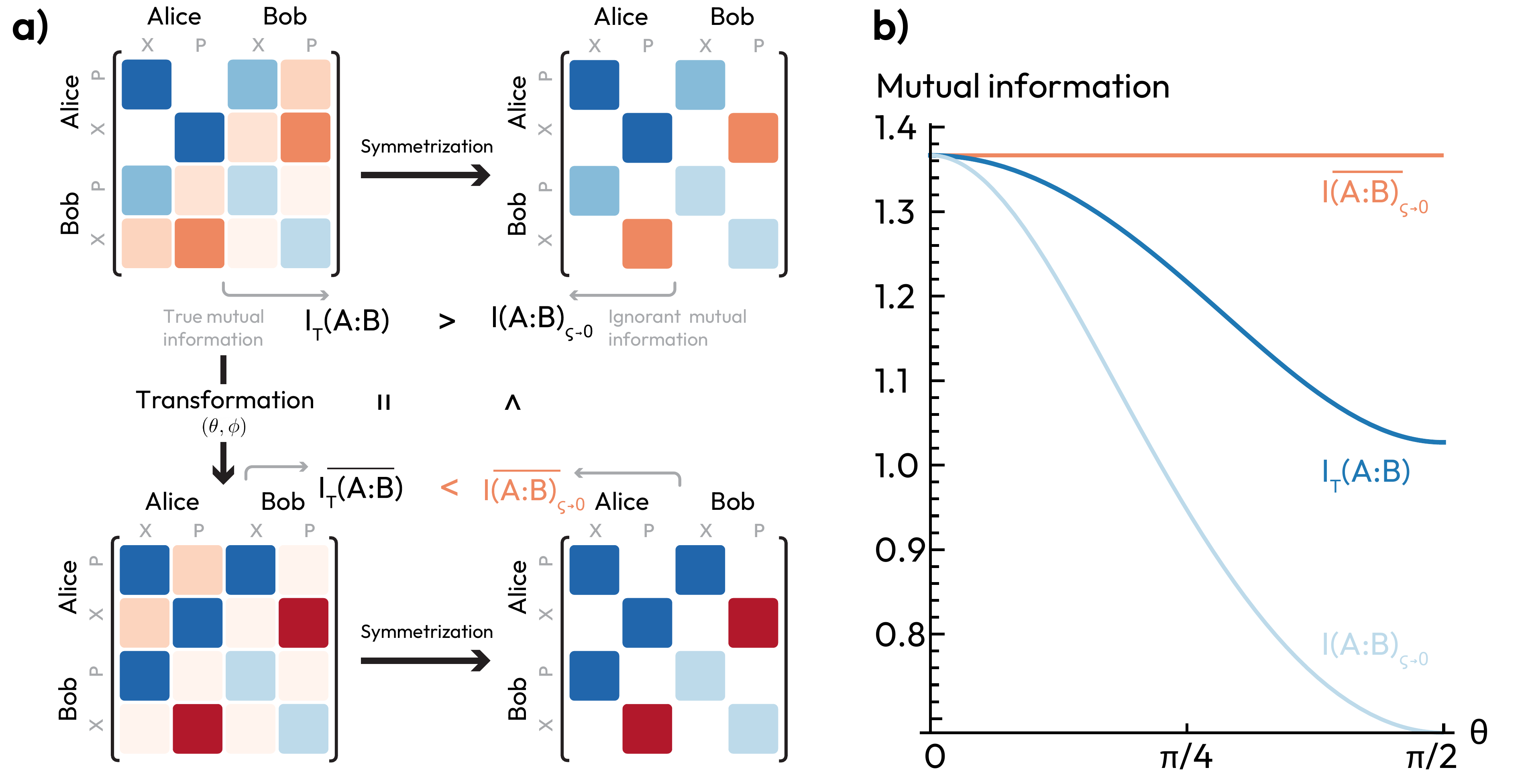}
        \caption{\textbf{a)} Illustration of different approaches for evaluation of the mutual information between Alice and Bob. Shade indicates the strength of the value, i.e. the darker the color the higher the absolute value of the matrix element. Symmetrization simplifies the security analysis at the cost of more conservative protocol performance. Data transformation aids the information reconciliation and improves the performance of the protocol. However, the post-transformation simplification leads to incorrect security assessment. \textbf{b)} Mutual information (in bits/channel use) dependency on a single misalignment angle $\theta$ ($\phi=0$) for different evaluation approaches. Symmetrization of the covariance matrix after data transformation $\overline{\Gamma}_{\varsigma\to 0}$ can lead to security overestimation.}
        \label{fig:illustration}
\end{figure*}

To reach the true MI, the quadratures cannot be regarded as independent any longer, and the error correction codes must be optimized accordingly. Alternatively, one could first transform the data such that the MI between Alice and Bob is the highest within the same quadratures, so that they can be treated as independent. Note that the true MI $I_T(A\!:\!B)$ in  Eq.(\ref{eq:mi-full}) will remain unchanged after the measurement regardless of any local linear transformation to the data. 

One can choose to transform Alice's or Bob's data sets to maximize the mutual information between individual quadratures $I(A:B)$. This can be achieved through a linear transformation of the form

\begin{equation}\label{eq:transform}
   \left[\begin{array}{c}
        \overline{x}  \\
        \overline{p}
   \end{array}\right]_{j}= \left[\begin{array}{cc}
       \text{cos}\,\Theta & \text{sin}\,\Theta \\
       \text{cos}\;\Phi & \text{sin}\;\Phi
   \end{array}\right]\left[\begin{array}{c}
        x  \\
        p 
   \end{array}\right]_{j},
\end{equation}
where $x(p)_{j}$ stand for sent/received data ($j=A, B$ for Alice and Bob respectively), and $\overline{x}(\overline{p})_{j}$ for transformed quadrature data on the respective side (here and further post-transformation values $\overline{\cdot}$ are denoted with an overline). The estimated parameters $\Theta$ and $\Phi$ differ depending on whether Alice or Bob performs the transformation. \par
Transforming Alice's modulation data aims to align her quadratures with those measured by Bob to maximize the mutual information. 
By definition $\langle x_a^2 \rangle = \langle p_a^2 \rangle$ and $\langle x_a p_a \rangle = 0$, hence this transformation does not alter the covariance matrix of Alice's data, focusing solely on maximizing the covariance between $\overline{x}(\overline{p})_a$ and $x(p)_a$. The transformation parameters \( \Theta = \tan^{-1}(\varsigma_{A^p,B^x}/\sigma^x_{A,B})\) and \( \Phi=\tan^{-1}(\varsigma_{A^x,B^p}/\sigma^p_{A,B}) \) maximize these covariances, effectively optimizing MI $I(A\!:\!B)_{\varsigma\to 0}$ between Alice and Bob.\par

Conversely, transforming Bob's data is more complex, since not only the covariance with between Alice's data Bob's data $\sigma^{x(p)}_{A,B}$ must be maximized, but also Bob's quadratures must remain independent $\varsigma_{B^x,B^p}\neq 0$.  Transformation (\ref{eq:transform}) modifies the variances of Bob's $x$-quadratures to \(\overline{V^x_B}= V^x_B \cos^2\Theta + V^p_B \sin^2\Theta + \varsigma_{B^x,B^p} \sin 2\Theta \) and Bob's $p$-quadratures to \(\overline{V^p_B}=  V^p_B \cos^2\Phi + V^x_B \sin^2\Phi + \varsigma_{B^x,B^p} \sin^2\Phi \).
The transformation parameters are now \( \Theta=\tan^{-1}(\varsigma_{A^x,B^p}/\sigma^x_{A,B})\) and \( \Phi=\tan^{-1}(\varsigma_{A^p,B^x}/\sigma^p_{A,B}) \) and maximize the covariances $\langle x_a\overline{x}_B\rangle$ and $\langle p_a\overline{p}_B\rangle$, but do not necessarily minimize the variances of Bob's transformed quadratures $\langle\overline{x}_B^2\rangle$ and $\langle\overline{p}_B^2\rangle$. The effectiveness of the transformation is contingent on the sign of \( \Theta \) and \( \Phi \) being opposite to that of \( \varsigma_{B^x,B^p} \). This constraint significantly limits when Bob's transformation can effectively increase the MI, making it a less flexible option compared to transforming Alice's data. 

\subsection{Modeling the imperfection}\label{sec:modelling}
In this section, we model the imperfection, specifically the presence of cross-correlation terms ($\varsigma\neq0$), as a consequence of Bob's unbalanced measurements.
A reference orthogonal phase space basis is needed to estimate the imbalance, hence we assume that the sender is able to modulate conjugate quadrature (each using Gaussian distribution with variance $V_A$), but Bob's heterodyne measurement has misalignment in both quadratures. Limited phase matching is the dominant effect leading to a non-orthogonal measurement basis \cite{hajomer2024long}, i.e. both quadratures of the received signal are phase-shifted from the original encoding basis. % resulting in each quadrature measurement to be subjected to separate and independent phase shift. 
We adopt the assumption of phase mismatch as the prevailing effect and proceed to model the cause of cross-correlations $\varsigma\neq 0$ (note that when used without any indices $\varsigma$ refers to all anti-diagonal elements in sub-matrices of $\Gamma$ in Eq. \ref{eq:cov-matrix}) as an independent phase shift $\theta$($\phi$) applied to respectively $x(p)$ quadrature.\par

The modulation strength of the quadratures can be modeled by re-scaling $\alpha$ of the random variables $\hat{x}_t$ and $\hat{p}_t$ that can be estimated from the covariance matrix $\Gamma$. For the more general case where both measured quadratures are misaligned (and symmetrical state modulation $V_A=V^x_A=V^p_A$), the elements of the covariance matrix (\ref{eq:cov-matrix}) are:
    \begin{align*}
    \sigma^x_{A,B}= \langle x_a\cdot x_B\rangle= \sqrt{\eta \tau_x}\cos[\theta] V_A \alpha,  \\  
    \varsigma_{A^x,B^p}= \langle x_a\cdot p_B\rangle= -\sqrt{\eta \tau_p}\sin[\phi] V_A \alpha,\\
    \varsigma_{A^p,B^x}=\langle p_a\cdot x_B\rangle= \sqrt{\eta \tau_x}\sin[\theta] V_A\alpha,   \\  
    \sigma^p_{A,B}=\langle p_a.p_B\rangle= -\sqrt{\eta \tau_p}\cos[\phi] V_A \alpha,\\
        V^x_B=\langle x_B\cdot x_B\rangle=1+\eta \tau_x(\alpha^2 V_A +\varepsilon),  \\ 
        V^p_B=\langle p_B\cdot p_B\rangle=1 +  \eta \tau_p( \alpha^2 V_A+\varepsilon), \\
        \varsigma_{B^x,B^p}=\langle x_B\cdot p_b\rangle= -\eta  \sqrt{\tau_x \tau_p} (\alpha^2 V_A+\epsilon) \sin[\phi + \theta],
    \end{align*}
where $\eta$ is the transmission of the channel, $\varepsilon$ is the channel noise, $\theta$ is the phase imbalance in x-quadrature, $\phi$ is the phase imbalance in p-quadrature, $\tau_x=\eta_D \eta_{bs}$, $\tau_p=\eta_D (1-\eta_{bs})$, $\eta_D$ is the efficiency of homodyne detection, $\eta_{bs}$ is the transmission of imbalanced beam splitter at the heterodyne detector. The balancing of the heterodyne beamsplitter transmission $\eta_{bs}$ is done only in the security model in order to eliminate differences between quadratures variances of the received state $V^{x(p)}_B$ that could remain after data normalization. We can estimate the phase imbalance ($\theta$, $\phi$), beam splitter imbalance $\eta_{bs}$ and the rescaling factor $\alpha$ from the covariance matrix $\Gamma$ as follows:
\begin{align}\label{eq:estimated}
    \theta&=\tan^{-1}\frac{\varsigma_{A^p,B^x}}{\sigma^x_{A,B}} \\ 
    \phi&=\tan^{-1}\frac{\varsigma_{A^x,B^p}}{\sigma^p_{A,B}} \\
    \eta_{bs}&=\frac{V^x_B-V^x_{B|A}}{\left(V^x_B-V^x_{B|A}\right)+ \left(V^p_B-V^p_{B|A}\right)} \\ 
    \alpha&=\frac{\sigma^x_{A,B}}{\sqrt{\eta \tau_x}\cos[\theta] V_m}=\frac{\sigma^p_{A,B}}{\sqrt{\eta \tau_p}\cos[\phi] V_m}
\end{align}
 We illustrate the issue in Fig. \ref{fig:illustration}(a) where the effect of symmetrization and transformation on the covariance matrix are shown, along with the MI derived from the respective matrix. After the transformation (\ref{eq:transform}) the ignorant MI improves $\overline{I(A\!:\!B)}_{\varsigma\to 0}\geq I(A\!:\!B)_{\varsigma\to 0}$ but can actually be higher than the true MI $\overline{I(A\!:\!B)}_{\varsigma\to 0}>\overline{I_T(A\!:\!B)}$ as it corresponds to MI of a perfectly implemented protocol. This might lead one to erroneously believe that modulated key data can be completely recovered, i.e. imperfections have no effect, but this is not accurate. Symmetrization of a post-transformation matrix $\overline{\Gamma}$ disregards the cross-correlation between the quadratures $\varsigma$ and results in an overestimation of a mutual information between Alice and Bob. Therefore, symmetrization is not allowed after the transformation has been applied. \par

    \section{Security analysis}\label{sec:keys}
Following the assumption that imperfect phase matching at the receiver side is the main cause of the strong cross-correlations $\varsigma$ and $\varsigma_{B^x,B^p}$ in Eq.(\ref{eq:cov-matrix}) we show in figure~(\ref{fig:illustration}b) how the MI diminishes rapidly with the phase shift. Even though the receiver obtains less information on one of the quadratures, it gains additional information regarding the conjugate one. The performance becomes equivalent to the coherent-state protocol with homodyne detection \cite{grosshans2003quantum}. \par
As long as Bob measures canonical conjugate quadratures (i.e. $\theta=-\phi$) all the MI lost due to phase misalignment can be recovered in post-processing using the transformation in Eq.~(\ref{eq:transform}). In the other case ($\theta\neq-\phi$) some amount of MI between Alice and Bob is irretrievably lost, which, as seen in Eq.~(\ref{eq:mut-snr}), for small values of noise $\varepsilon_{x(p)}$ can be approximated to 

    \begin{equation} \label{eq:lost-MI}
        \begin{split}
    & I(B^x\!:\!B^p) =\log_2\left[\eta\left(V_m+\varepsilon+\frac{2}{\eta^2}\right)\right]- \\ & \frac{1}{2}\log_2\left[\eta^2\!\left(V_m+\varepsilon+\frac{2}{\eta^2}\right)^2 + \left(\eta (\varepsilon + V_m) \sin[\phi + \theta]\right)^2\right]\ .
        \end{split}
    \end{equation}
The effect of the transformation depends on precision of estimation of misalignment angles $\theta$ and $\phi$, hence will lead to limited MI recovery in practice. 
%%%%
\subsection{Security in asymptotic regime}
Similar to the MI which can be evaluated with and without symmetrization, the upper bound on the accessible information $\chi(E)$ of Eve can be evaluated with or without. From one perspective, the symmetrization of the covariance matrix increases $\chi(E)$, as a consequence of erroneously identifying part of the signal from the other quadrature as noise.  
Conversely, avoiding symmetrization and admitting a more comprehensive description of the setup allows to limit our estimation on the upper bound of information accessible to Eve at the cost of a more exhaustive analysis and processing of the experimental data. 
Careful characterization of the experimental setup can identify and quantify the dominant imperfection and its effects.\par
To recapitulate, there are two approaches for security analysis with different levels of involvement of experimental data analysis. One approach makes more pessimistic assumptions in the security analysis than the other. Consequently, each approach introduces a distinct evaluation method for mutual information and for the Holevo bound, summarized in Tab.(\ref{tab:MI-chi}). 

    \begin{table}[]
    \centering
\begin{tabular}{l|l|l}
                            & $I_T(A:B)$ & $I(A:B)_{\varsigma\to 0}$ \\  \hline
$\chi_T(E)$                 &   $K_{TT}$    &    $K_{IT}$     \\      \hline
$\chi(E)_{\varsigma\to 0}$  &     $K_{TI}$    &     $K_{II}$    
\end{tabular}
\caption{Approaches to evaluation of secret key rate components. Alice and Bob can choose to evaluate their mutual information and/or Eve's information based on full $\Gamma$ or simplified $\Gamma_{\varsigma\to 0}$ covariance matrix leading to a four possible secure key rates.}\label{tab:MI-chi}
    \end{table}

\begin{figure}
        \centering
    \includegraphics[width=.99\linewidth]{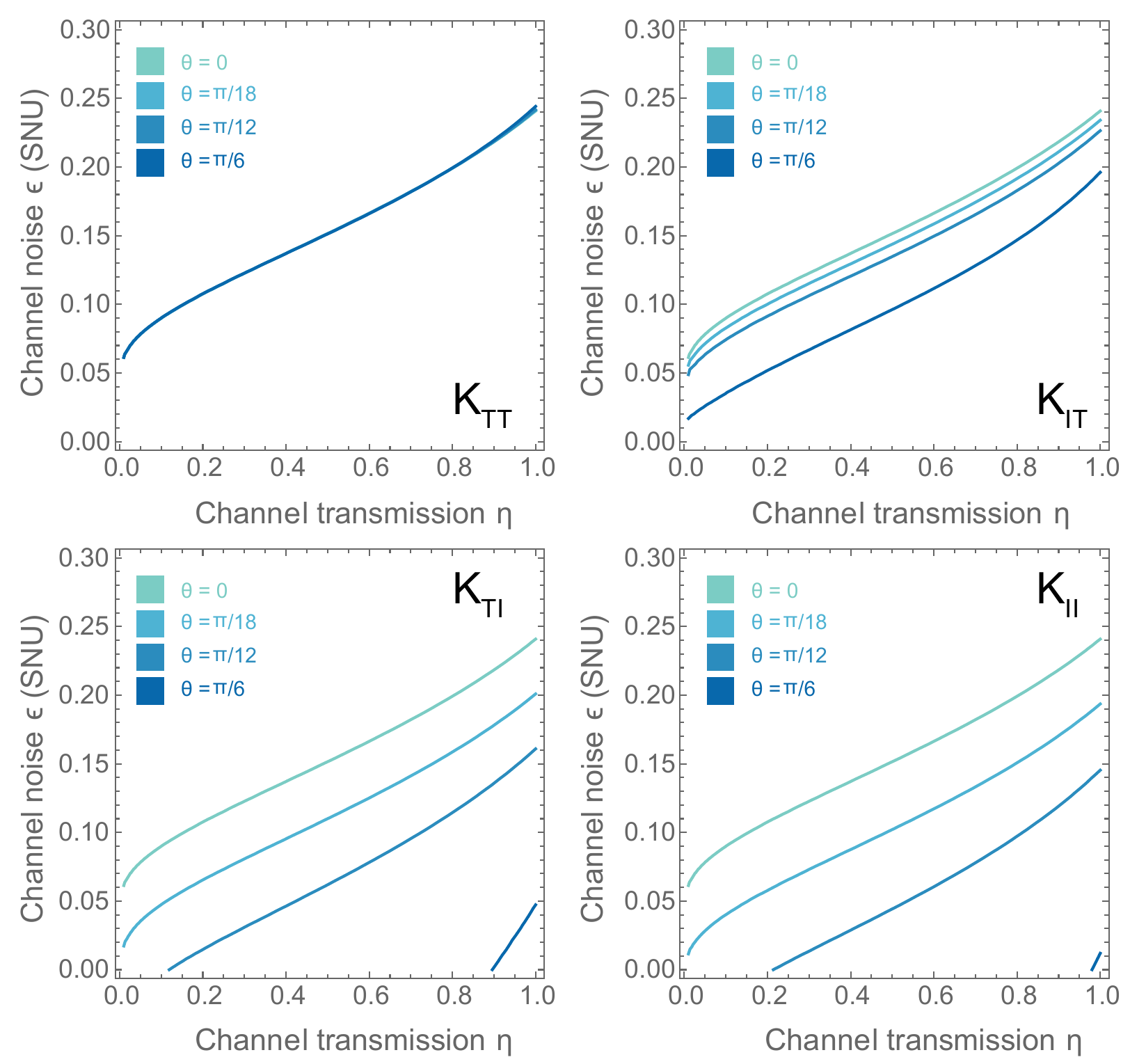}
        \caption{Maximal tolerable excess noise $\varepsilon$  dependence on the channel loss $\eta$ for different models (summarized in Tab.~(\ref{tab:MI-chi})) under various misalignment angles $\theta=0,\,\pi/18,\,\pi/12,\,\pi/6$ in asymptotic regime. Reconciliation efficiency $\beta=95\%$, modulation variance $V_m=3.3$ SNU.}
        \label{fig:4-grid}
\end{figure}

In Fig.~\ref{fig:4-grid} we analyze the maximal tolerable channel loss $\eta$ and excess noise $\varepsilon$ for each theoretical model in the asymptotic regime. The ability to reconcile the information in both quadratures and correct the results ($K_{TT}$) will allow to tolerate larger noise and loss values with any phase mismatch value. Ignoring correlations only when evaluating MI between Alice and Bob, but using the general covariance matrix $\Gamma$ for the Holevo ($K_{IT}$) decreases the noise and loss tolerance, and the performance of the protocol. Adopting the Holevo bound $\chi(E)){\varsigma\to 0}$ from the simplified covariance matrix (i.e. key rates $K_{TI}$ and $K_{II}$) significantly limits the range of secure channel properties. Even minor phase deviations lead to a substantial key rate decrease, and may lead to a security break already in a noiseless channel which is in line with previously shown effective decrease of estimated channel transmittance $\eta$ and increase of noise $\varepsilon$ by the phase noise \cite{jouguet2012analysis}. Due to limited performance of the latter two approaches we focus on the former two ($K_{TT}$ and $K_{IT}$) for security analysis of the experimental data. 

\subsection{Finite-size effects}\label{sec:finites}

In practice only a finite amount of signals can be exchanged between trusted parties which imposes limitations on the accuracy of parameter estimation and must be taken into account during security analysis, in the presence of imbalance in the heterodyne measurement we are required to estimate the phase imbalance as well on top of the channel parameters. We have established in the previous section that the security of the protocol degrades with the degree of imbalance, so we consider an upper bound for the estimated imbalance $\delta=\theta+\phi$.\par
It has been demonstrated that conducting error correction prior to parameter estimation enables the utilization of all available measurements for both key generation and error correction \cite{leverrier2015composable}. However, achieving the true $I_{T}(A:B)$, is not feasible because Alice lacks the necessary information about the imbalance required to execute the transformation for retrieving the genuine mutual information. Hence to enable the transformation (\ref{eq:transform}) it is necessary to perform parameter estimation, at least for estimating the total imbalance $\delta$, prior to error correction. Consequently, we examine two potential finite key rates:

 \begin{equation}
    K_{\{n\}}=\frac{n}{N}\left[K^{\infty}_{TT}\left(t_\text{low}, \varepsilon_\text{up},\delta_{up}\right)-\Delta(n)\right],
    \end{equation}
\begin{equation}
    K_{\{N\}}=K^{\infty}_{IT}\left(t_\text{low}, \varepsilon_\text{up},\delta_{up}\right)-\Delta(N)
    \end{equation}

% Consequently, the key rate in the finite-size regime becomes dependent on the block size $n$ used for the key:
%     \begin{equation}\label{eq:finite-key}
%         K(n)=\frac{n}{N}\left[K\left(\eta_\text{low}, \varepsilon_\text{up},\delta_\text{up}\right)-\Delta(n)\right],
%     \end{equation}
where asymptotic key rate $K$ (\ref{eq:skr-asymptotic}) is chosen according to one of the approaches detailed in Tab.~\ref{tab:MI-chi}, estimated worst-case channel parameters $\varepsilon_\text{up}$, $\eta_\text{low}$ and the largest imbalance $\delta=\theta+\phi$. The lower bound on the key in such regime is now confined by $\Delta(n)$, which represents the reduction of the rate due to limited number of exchanged signals $n$ used for key formation, which is necessarily lower than the overall amount of signals received and measured $n<N$ \cite{leverrier2010finite, leverrier2011continuous}. Signals used for parameter estimation of $\varepsilon$, $\eta$ and $\delta=\theta+\phi$ are discarded and not used for key generation. Estimates of are taken with Gaussian confidence intervals corresponding to an error probability of $10^{-10}$ as: $\mathbf{E}(\delta_\text{up})=\delta+6.5\sqrt{\text{Var}\left(\delta\right)}$, $\mathbf{E}(\eta_\text{low})=\eta-6.5\sqrt{\text{Var}(\eta)}$ and  $\mathbf{E}(\varepsilon_\text{up})=\varepsilon+6.5\sqrt{\text{Var}(\varepsilon)}$, see the Supplementary Materials for derivation of $\text{Var}(\delta)$,$\text{Var}(\eta)$ and $\text{Var}(\varepsilon)$.\par
Note that any possible heterodyne imbalance, characterized by $\eta_{bs}$, is assumed to be fixed for the duration of key generation and assumed to be under full control of Bob, and can thus be determined precisely. Hence, the presented parameter estimation allows to compute confidence intervals of all relevant parameters, thus making the analysis compatible with the composable security framework \cite{pirandola2021composable, pirandola2024improved}. 

%%%%%%%%%%%%%%%%%%%%%%%%%%%%
    \section{Experimental validation } \label{chapter:experiment}

    \begin{figure*}[ht]
        %\centering
    \includegraphics[width=.99\linewidth]{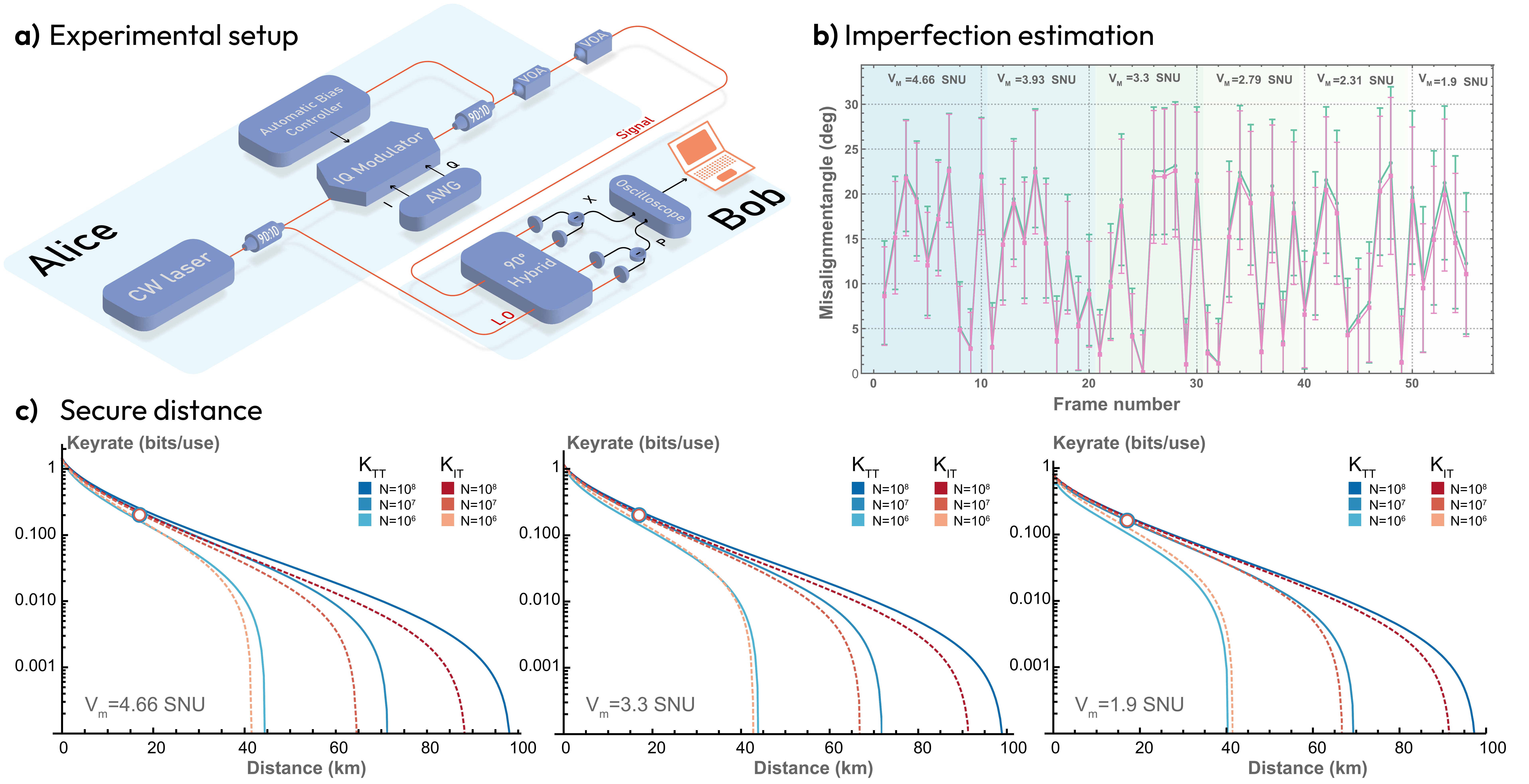}
        \caption{\textbf{a)} Experimental setup for imbalance characterization. Alice sends an ensemble of coherent states generated by modulated 1550 nm continuous wave (CW) laser using IQ modulator driven by an arbitrary waveform generator (AWG). To operate the IQ modulator in carrier suppression mode, an automatic bias controller is used.  Variable optical attenuators are used to control the variance of the modulated state and simulated fiber loss. To measure the orthogonal quadrature, a tunable $90^{\circ}$ optical hybrid plus two balanced detectors are deployed. \textbf{b)} To test if our model fits the experimental data we check if the misalignment angle calculated from correlations between Alice and Bob's data: $\tan^{-1}\left[\varsigma_{A^p,B^x}/\sigma^x_{A,B}\right]+\tan^{-1}\left[\varsigma_{A^x,B^p}/\sigma^p_{A,B}\right]$ (green) is the same as the misalignment angle calculated from correlations between Bob's quadratures $\sin^{-1}\left[\varsigma_{B^x,B^p}/\sqrt{(V^x_B-1)(V^p_B-1)}\right]$ (pink). The mean value if misalignment $\theta+\phi=10^{\circ}$ is indicated by the dashed line.
        \textbf{c)} Secure key rate in finite-size regime dependency on the distance over standard fiber (0.2dB/km) for different block sizes $N=10^6,\,10^7,\,10^8$. Blue (solid) lines show the security analysis approach based on transformation $K_{TT}$ with optimized parameter estimation data fraction; Red (dashed) lines show the simplified security analysis without transformation. Points (at 17km distance) correspond to experimantally achieved values. Modulation variance $V_m$ corresponds to the experimental values (see part b), with averaged values (over all frames) of estimated excess noise $\varepsilon=5\cdot 10^{-3}$ SNU and mean misalignment value $\theta+\phi=10$ deg.}
        \label{fig:experiment}
    \end{figure*}
To test the validity of the theoretical method for performance improvement we conduct a set experimental measurements and consequent parameter estimation and security analysis. 
Figure \ref{fig:experiment}a shows a simplified scheme of the  prepare-and-measure CVQKD setup used for characterizing the effect of imperfect heterodyne measurements due to the phase and amplitude imbalance of the LO in CVQKD receivers. As a first step, Alice digitally prepared displaced coherent states with quadrature components $x$ and $p$ drawn at a rate of 20 Mbaud from two independent and identical Gaussian distributions. The quantum symbols were then upsampled to 1 GSample/s and pulse-shaped using a root-raised cosine (RRC) anti-aliasing filter with a roll of 0.2. To avoid the low-frequency noise both at the transmitter and the receiver, the quantum signal was upconverted to $\omega/2\pi = 60$ MHz, i.e multiplied by $\cos \omega t$, for  double sideband modulation. Finally, Alice uploaded the prepared waveform into an arbitrary waveform generator (AWG) with a sampling rate of 1 Gsample/s and a vertical resolution of 16 bits.
The optical setup was built from polarization maintaining fiber components. At Alice's station, a continuous wave (CW) laser operating at 1550 nm was used as an optical carrier and shared with Bob as a LO to avoid the laser phase noise in the local LO (LLO) CVQKD scheme \cite{qi2015generating}. An in-phase and quadrature (IQ) modulator driven by the AWG was used to encode an ensemble of coherent states into the optical carrier. The bias voltages to the IQ modulator were controlled using an automatic bias controller. The modulated optical signal was suitably attenuated using a variable optical attenuator (VOA) so that the quantum band contained few photons. The quantum signal was then transmitted through a quantum channel emulated by an VOA with a loss of 3.5 dB corresponding to a 17 km optical fiber channel (at the loss of 0.2 dB/km). 

At Bob's station, the phase and amplitude quadratures of the quantum signal were measured using a $90^{\circ}$ optical hybrid, which is an optical circuit splitting the quantum signal and the LO into two arms and mixing them in 3 dB beamsplitters with a phase difference of $90^{\circ}$ for the LO arms. In this circuit, the phase difference between the LOs arms can be tuned by applying a voltage on the hybrid. For our purpose of characterizing the effect of imperfect measurements, the phase difference between the LO arms was tuned within $\approx 90^{\circ}\pm 10^{\circ}$. The measured quadratures were detected using two
homemade broadband balanced detectors each with a bandwidth of 250 MHz and a quantum efficiency of $\eta_D\approx 85\% $. The output of the balanced detectors was sampled using an oscilloscope with a sampling rate of 1 GSample/s, which was synchronized with the AWG using a 10 MHz clock reference. Afterwards, the quantum symbols were recovered by applying an offline digital signal processing pipeline including baseband transformation, reference symbols aided time synchronization, RRC matched filter, and downsampling.  
The measurement time was divided into frames each consisting of $10^6$ Gaussian distributed complex values. The measurement was taken over a range of modulation variance $V_{m} $ from 1.6 to 4.5 shot noise units (SNU) with 10 frames for each $V_{m} $ value. Finally, we estimated the imperfection of Bob's measurement from the data and executed a transformation on Alice's dataset as described in Sec. \ref{sec:transformation}. 

Note that statistical variation of the phase shifts $\theta$ and $\phi$ have significant impact on the feasibility of secure key distribution \cite{jouguet2012analysis}, thus we verify that in our experiment we can resolve phase with sufficient precision and make sure it is not fluctuating during the measurement (see Supplementary Materials for further details). 

Results of a misalignment angle estimation (as described in Sec.2\ref{sec:modelling}) in the experiment are shown in Fig.\ref{fig:experiment}b, where we check if the misalignment angles calculated from correlations between Alice and Bob's data as in Eq.~(\ref{eq:estimated}) are the same as the one calculated from the correlations between Bob's quadratures $\sin^{-1}\left[\varsigma_{B^x,B^p}/\sqrt{(V^x_B-1)(V^p_B-1)}\right]$.

\subsection*{Performance of the system}

Based on theoretical analysis in Sec. \ref{sec:keys} we choose two approaches that allows to achieve the highest secure key rate: $K_{TT}$ that requires phase misalignment $\delta$ estimation before error correction, and $K_{IT}$ that avoids phase misalignment $\delta$ estimation before error correction at the cost of more pessimistic value of MI $I(A:B)_{\varsigma\to 0}$. The former approach is expected to reach higher secure key rates, but necessary parameter estimation implies some data must be discarded which will translate into larger finite-size effects. Avoiding the transformation with $K_{IT}$ allows to perform error correction prior to channel estimation \cite{leverrier2015composable}, which incurs lower finite-size effects penalty, despite leading to lower key rate in asymptotic regime. 
Figure (\ref{fig:experiment}c) combines all the previous analysis and compares the secure key rates with finite-size effects (see Sec.3\ref{sec:finites}) under different models for different values of measured excess noise $\varepsilon$ and misalignment angles $\delta$, as well as the distance to which the system could be securely used under an assumption that respective parameters remain fixed. Under measured misalignment approaches $K_{TI}$ and $K_{II}$ could not establish a positive key in asymptotic regime. To reach the best performance not only additional estimation and data transformation is required, but crucially an optimization of a fraction of data used for the secure key $n/N$. Without the latter optimization the key rate advantage (comparing to $K_{IT}$ approach) is not fully extracted in the finite-size regime. 

    \section{Discussion}
CVQKD presents a viable option for large-scale deployment and integration into existing telecommunication networks, as it utilizes standard coherent detection methods—homodyne and heterodyne detection—that operate reliably at room temperature. However, practical implementations of coherent detection in CV QKD often suffer from imperfections, some of which can be ignored at the cost of limited system performance. In this work, we introduced a finite-size security proof that accounts for the phase imbalance effects arising from imperfect heterodyne measurements in practical CVQKD systems, aiming to enhance system performance by refining security models. A major contributor to these imperfections is the phase imbalance inherent in QKD protocol implementations, particularly pronounced in photonic integrated phase-diversity receiver ~\cite{hajomer2024continuous} due to manufacturing tolerances or thermal drift and phase-locking challenges in bulk receivers. 

We demonstrated that phase imbalances induce cross-correlations and contaminate the data exchanged between Alice and Bob. This contamination not only reduces the mutual information available to trusted parties but also allows an eavesdropper to maintain a relative information advantage, thereby diminishing the achievable key rate and restricting the acceptable phase imbalance range in the receiver.

In re-evaluating the security-related quantities, we identified possible approaches for authentic experimental operation. A straightforward and commonly used method relies on symmetrization, which does not require additional analysis or processing but leaves no room for theoretical optimization; the protocol’s feasibility is then dependent on experimental advancements. We propose an alternative compensation method that employs a local transformation of the received quantum symbols, thereby restoring mutual information between Alice's and Bob's corresponding quadratures without underestimating the eavesdropper’s information. Assessment of accessible information of an eavesdropper without resorting to symmetrization allows to improve the performance of the system in presence of the phase misalignment. When extended to evaluation of mutual information, such comprehensive approach can further promote the secure operation of a QKD system. 

We validated our approach by experimentally testing a heterodyne-based system with induced phase imbalance, evaluating parameter estimation techniques, and confirming the effectiveness of our theoretical approach to recover system performance. Additionally, we extend the finite-size analysis to rigorously bound phase imbalance imperfections in heterodyne detection under stricter conditions, making our method directly applicable to practical CV QKD systems. Our analysis of finite-size effects reveals that, for longer distances, performing parameter estimation prior to error correction is more advantageous for achieving higher key rates, given the necessity of imbalance estimation for implementing our transformation. In contrast, for shorter distances, higher key rates are achieved when error correction precedes parameter estimation.

Our findings highlight the critical importance of accounting for cross-correlations under imperfect heterodyne measurement and suggest practical strategies for enhancing CVQKD performance while relaxing the phase imbalance requirement on photonic-integrated CVQKD receivers.

    \section*{Acknowledgments}
    This project was funded within the QuantERA II Programme (project CVSTAR) that has received funding from the European Union’s Horizon 2020 research and innovation program (Grant Agreement No.  101017733, 8C22002),  and the European Union’s Horizon Europe research and innovation programme under the project ”Quantum Security Networks Partnership” (QSNP, grant agreement No. 101114043). A. O. acknowledges the project IGA-PrF-2024-008 of Palacky University Olomouc. I.D. acknowledges support from the project 22-28254O of the Czech Science Foundation. V.C.U. acknowledges support from the project 21-44815L of the Czech Science Foundation and the project CZ.02.01.01/00/22\_008/0004649 (QUEENTEC) of the Czech MEYS. A. H. and T.G. acknowledge funding from the Carlsberg Foundation project CF21-0466. A.H., U.L.A., T.G. acknowledge support support from Innovation Fund Denmark (CyberQ, grant agreement no. 3200-00035B)
    \section*{Data availability statement}
    The data that support the findings of this study are available upon reasonable request from the authors.
\bibliography{lib}

%%%%%%%%%%%%%%%%%%%%%%%%%%%%%%%%%%%%%%%%%%%%%%%%%%%%%%%%%
%%%%%%%%%% Merge with supplemental materials %%%%%%%%%%
\pagebreak
%\widetext
\onecolumn
\begin{center}
\textbf{\large Supplemental Materials: Imperfect heterodyne measurement in continuous variable quantum key distribution}
\end{center}
%%%%%%%%%% Prefix a "S" to all equations, figures, tables and reset the counter %%%%%%%%%%
\setcounter{equation}{0}
\setcounter{section}{0}
\setcounter{figure}{0}
\setcounter{table}{0}
\setcounter{page}{1}
\makeatletter
\renewcommand{\theequation}{S\arabic{equation}}
\renewcommand{\thefigure}{S\arabic{figure}}
\renewcommand{\bibnumfmt}[1]{[S#1]}
\renewcommand{\citenumfont}[1]{S#1}
\section{Mutual information}\label{sec:mutual}
The mutual information between Alice and Bob is:
\begin{equation}
    I_{AB}=H(x_b,p_b)-H(x_b,p_b|x_m,p_m)
\end{equation}
Where $H(x_b,p_b)$ and $H(x_b,p_b|x_m,p_m)$ are the entropies of the bi-variant Gaussian distributions with covariance matrices $\gamma_B$ and $\gamma_{B|A}$. 
\begin{equation}
\begin{split}
    H(x_b, p_b) &= -\int \int \mathcal{N}(0, \gamma_B) \log[\mathcal{N}(0, \gamma_B)] \, dx_b \, dp_b \\
    &= -\mathbb{E} \left[\log[\mathcal{N}(0, \gamma_B)]\right] \\
    &= -\mathbb{E} \left[\log \left(\frac{1}{2 \pi \sqrt{|\gamma_B|}} \exp\left(-\frac{1}{2} X_B^T \gamma_B^{-1} X_B\right)\right)\right] \\
    &= \log[2 \pi \sqrt{|\gamma_B|}] + \frac{1}{2} \mathbb{E}\left[X_B^T \gamma_B^{-1} X_B\right]    
\end{split}
\end{equation}

Where $X^T_B=\{x_b,p_b\}$
\begin{equation}
    \mathbb{E}[X^T_B \gamma^{-1}_B X_B]=\frac{1}{V^x_B V^p_B-C^2_{xp}}\mathbb{E}\left[\left[\begin{array}{cc}
       x_b  &  p_b
    \end{array}\right]\left[\begin{array}{cc}
     V^x_B    & -\varsigma_{B^x,B^p} \\
       -\varsigma_{B^x,B^p}  & V^p_B
    \end{array}\right]\left[\begin{array}{c}
       x_b  \\  p_b
    \end{array}\right]\right]=2
\end{equation}
So, we have 
\begin{equation}
    H(x_b,p_b)=\log[2\pi\sqrt{|\gamma_B|}]+1
\end{equation}
similarly
\begin{equation}
    H(x_b,p_b|x_m,p_m)=\log[2\pi\sqrt{|\gamma_{B|A}|}]+1
\end{equation}
Therefore
\begin{equation}
    I_{AB}=\frac{1}{2}\log\left[\frac{|\gamma_B|}{|\gamma_{B|A}|}\right]=\frac{1}{2}\log\left[\frac{|\gamma_A|}{|\gamma_{A|B}|}\right]
\end{equation}
The mutual information can be expressed in terms of the $SNR$ between Alice and Bob, as well as the mutual information between Bobs quadratures. 
\begin{equation}
\begin{split}
    I_{AB} &= \frac{1}{2} \log\left[\frac{|\gamma_B|}{|\gamma_{B|A}|}\right] = \frac{1}{2} \log\left[\frac{V^x_B V^p_B - C_{xp}^2}{V^x_{B|A} V^p_{B|A} - C_{B|A}^2}\right] = \frac{1}{2} \log\left[\frac{V^x_B V^p_B - C_{xp}^2}{V^x_{B|A} V^p_{B|A} - C_{B|A}^2}\right] \\
    &= \frac{1}{2} \log\left[V^x_B V^p_B \left(\frac{V^p_B - C_{xp}^2 / V^x_B}{V^p_B}\right)\right] - \frac{1}{2} \log\left[V^x_{B|A} V^p_{B|A} \left(\frac{V^p_{B|A} - C_{B|A}^2 / V^x_{B|A}}{V^p_{B|A}}\right)\right] \\
    &= \frac{1}{2} \left(\log\left[V^x_B V^p_B\right] - \log\left[V^x_{B|A} V^p_{B|A}\right]\right) - I(x_b; p_b) + I(x_{b|a}; p_{b|a}) \\
    &=\frac{1}{2}\left(\log[(1+\eta \tau_x[\alpha^2 V_A +\varepsilon]) (1+\eta \tau_p[\alpha^2 V_A +\varepsilon])]-\log[(1+\eta \tau_x\varepsilon) (1+\eta\tau_p\varepsilon)]\right)- I(x_b; p_b) + I(x_{b|a}; p_{b|a})\\
    &= \log_2(1 + SNR) - I(x_b : p_b) + I(x_{b|a} : p_{b|a})
\end{split}
\end{equation}
When the $SNR$ is precisely estimated, it remains unaffected by imbalance. Therefore, the aforementioned expression indicates that despite the $SNR$ being independent of imbalance, neglecting the correlation between Bob's quadrature will result in an overestimation of both the mutual information and, consequently, the key rate shared by Alice and Bob.
\section{Excess noise}

For imbalance heterodyne measurement, the quadratures can not be considered independent of each other while estimating the excess noise. In Ref. \cite{Xiang2023} the authors consider the quadratures separately while calculating the excess noise which lead to overestimation of excess noise. They claim that the local transformations on the Bob's data completely negate any negative impacts of the imperfect heterodyne measurements. We show, on the contrary, that local transformation on Bob's or Alice's data does not compensate for the imperfections in the measurement. The true mutual information, excess noise and the Holevo bound when estimated properly are independent of any local transformations on the data and the key rate is always less than the key rate for perfect implementation of the protocol as long as the quadratures measured are not canonical conjugates. We obtain the excess noise as 

    \begin{equation}
    \mathbf{V}_{\varepsilon}=V_{B|A}-I,
    \end{equation}
    
where $V_{B|A}=V_B-C_{AB}.V^{-1}_A.C^T_{AB}$.

The channel noise before hetrodyne measurement can also be obtained from the correlation between the Bob's quadratures as follows:
\begin{equation}\label{eq:true-noise}
    \eta \epsilon=\frac{\varsigma_{B^x,B^p}-V^{1,2}_{B|A}}{\sqrt{\tau_x\tau_p}Sin[\phi+\theta]}
\end{equation}
Note that Eq.\ref{eq:true-noise} is always non-negative value and indicates an authentic level of observable excess noise in the practical system, although the actual value adopted for the security analysis will always be larger due to security concerns. 

\section{Phase shift fluctuations}
\begin{figure}
    \centering
    \includegraphics[width=0.5\linewidth]{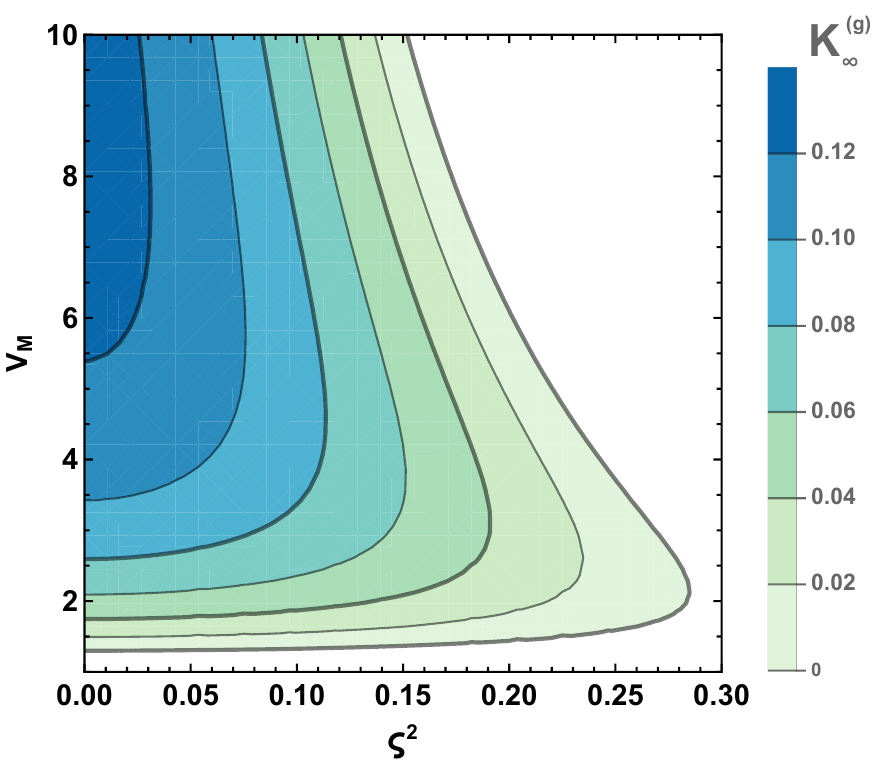}
    \caption{Effect of normal phase shift fluctuation on the optimality of modulation variance $V_m$ (in SNU). Postprocessing efficiency $\beta=95\%$, channel loss $\eta=5$ dB, and excess noise $\varepsilon=1\%$ SNU.}
    \label{fig:fluctuations}
\end{figure}

Statistical variation of the phase shift $\theta$ can significantly impact the range of secure parameters and feasibility of secure key distribution in the first place even in asymptotic regime \cite{jouguet2012analysis}. Assuming its distribution is zero centered Gaussian $p(\theta)=\mathcal{N}(0,\varsigma^2)$, the secret key rate now strongly depends on its variance as $\langle\cos\theta\rangle=\exp{\left[-\varsigma^4/4\right]}$. Phase fluctuations impose additional restrictions towards the choice of optimal modulation variance $V_m$ (see Fig. \ref{fig:fluctuations}), and crucially determine the resolution required for phase estimation which should be lower than $\varsigma^2<10^{-2}$ to ensure the optimal performance of the CV QKD protocol. In the current work we ensure that fluctuations of estimated phase shifts $\theta$ and $\phi$ for the duration of a single frame are negligible and thus treated as fixed. 

\section{Using the imperfection to normalize the data (Optional)}
To estimate the excess noise, we must first normalize the data with respect to shot noise. The shot noise is influenced by the detector's responsivity, which may vary over time for various reasons. Hence shot noise needs to be estimated separately each time a key needs to be generated. For the case of imbalance hetrodyne measurement, we can instead use the correlations between Bob's quadrature $\varsigma_{B^x,B^p}$ to estimate the shot noise.
\begin{equation}
    \frac{\sqrt{\tau_x}\varsigma_{B^x,B^p}}{\sqrt{\tau_p} Sin[\phi+\theta]}=\eta \tau_x(\alpha^2 V_m+\epsilon)
\end{equation}
\begin{equation}
    \frac{\sqrt{\tau_p}\varsigma_{B^x,B^p}}{\sqrt{\tau_x} Sin[\phi+\theta]}=\eta \tau_p(\alpha^2 V_m+\epsilon)
\end{equation}
So the normalization factor is
\begin{equation}
    V^{x(p)}_N=V^{x(p)}_B-\frac{\sqrt{\tau_{x(p)}}\varsigma_{B^x,B^p}}{\sqrt{\tau_{p(x)}} Sin[\phi+\theta]}-V_{elec}
\end{equation}
Where $V_{elec}$ is the electronic noise of the detectors and $V^{x(p)}_B$ is the unnormalized variances of Bob's quadratures.  
\section{Finite-size effects}

In practice only a finite amount of signals can be exchanged between trusted parties which imposes limitations on the accuracy of parameter estimation and must be taken into account during security analysis. In the presence of imbalance in the heterodyne measurement we are required to estimate the phase imbalance as well on top of the channel parameters. We have established in the previous section that the security of the protocol degrades with the degree of imbalance, so we consider upper bound for the estimated imbalance $\delta=\theta+\phi$, the finite size key rate for the pessimistic parameters corresponding to the failure probability of $10^{-10}$ is: 

    \begin{equation}
    K_{\{n\}}=\frac{n}{N}\left[K_\infty\left(t_\text{low}, \varepsilon_\text{up},\delta_{up}\right)-\Delta(n)\right].
    \end{equation}

The lower bound on the key in such regime is now confined by $\Delta(n)$, which represents the reduction of the rate due to limited number of exchanged signals $n$ used for key formation, which is necessarily lower than the overall amount of signals received and measured $n<N$ \cite{leverrier2010finite, leverrier2011continuous}.

%\lipsum[3]%placeholder
Rest of the signals are used estimate the channel parameters and the imbalance in the measurement. 
We start by defining the estimator of the imbalance as 

    \begin{equation}
    \widehat{\mathbf{T}}^{\theta}= \frac{\frac{1}{m}\sum^m_{i=1}M^p_iB^x_i}{\frac{1}{m}\sum^m_{i=1}M^x_iB^x_i}
    \end{equation}
    
    \begin{equation}
    \widehat{\mathbf{T}}^{\phi}= \frac{\frac{1}{m}\sum^m_{i=1}M^x_iB^p_i}{\frac{1}{m}\sum^m_{i=1}M^p_iB^p_i}
    \end{equation}

\begin{equation}
    \mathbb{E}[\widehat{\mathbf{T}}^{\theta}]= \mathbb{E}\left[\frac{\frac{1}{m}\sum^m_{i=1}M^p_iB^x_i}{\frac{1}{m}\sum^m_{i=1}M^x_iB^x_i}\right]=\mathbb{E}\left[\frac{\mathbf{x}}{\mathbf{y}}\right]
\end{equation}
Where $\mathbf{x}$ and $\mathbf{y}$ defined as

\begin{equation}
     \mathbf{x}:=\frac{1}{m\sqrt{\eta \tau_x}V_m \alpha Cos\theta } \sum^m_{i=1}M^p_iB^x_i,\hspace{0.3cm}\mathbf{y}:=\frac{1}{m\sqrt{\eta \tau_x}V_m \alpha Cos\theta } \sum^m_{i=1}M^x_iB^x_i
\end{equation}
If large number of states are used for parameter estimation, $Var[\mathbf{y}]<<1$. We can reasonably make the approximation $1/\mathbf{y}\approx 2-\mathbf{y}$ by ignoring higher order terms from the Taylor series expansion of $1/\mathbf{y}$ around 1, as $\mathbb{E}[\mathbf{y}]=1$. So, we have 
\begin{equation}
    \mathbb{E}[\widehat{\mathbf{T}}^{\theta}] =\mathbb{E}\left[\frac{\mathbf{x}}{\mathbf{y}}\right]\approx \mathbb{E}[\mathbf{x}(2-\mathbf{y})]=2 \mathbb{E}[\mathbf{x}]-\mathbb{E}[\mathbf{x}\mathbf{y}]=Tan\theta+\frac{Tan\theta}{m}-\frac{1}{m \eta \tau_x V^2_t \alpha^2 Cos^2\theta}\mathbb{E}[p_m x_b x_m x_b]
\end{equation}
Fluctuating transmission or imbalance might introduces correlations between random variables $\mathbf{x}$ and $\mathbf{y}$. In the absence of these fluctuations $\mathbb{E}[\mathbf{x}\mathbf{y}]=\mathbb{E}[\mathbf{x}]\mathbb{E}[\mathbf{y}]$ and hence $\mathbb{E}[\widehat{\mathbf{T}}^{\theta}]=Tan\theta$. 
\begin{equation}
    Var[\widehat{\mathbf{T}}^{\theta}]=Var\left[\frac{\mathbf{x}}{\mathbf{y}}\right]
\end{equation}

\begin{equation}
   Var[\widehat{\mathbf{T}}^{\theta}] =Var\left[\frac{\mathbf{x}}{\mathbf{y}}\right]\approx Var[\mathbf{x}(2-\mathbf{y})] = 4Var[\mathbf{x}]+Var[\mathbf{x}\mathbf{y}]-4 (\mathbb{E}[\mathbf{x}^2\mathbf{y}]-\mathbb{E}[\mathbf{x}]\mathbb{E}[\mathbf{x}\mathbf{y}] )
\end{equation}

\begin{equation}
\begin{split}
    \mathbb{E}[\mathbf{x}^2 \mathbf{y}] &= \frac{1}{m^3 (\eta \tau_x)^{3/2} V_t^3 \alpha^3 \cos^3 \theta} \, \mathbb{E}\left[\sum_{i=1}^m M^p_i B^x_i \sum_{i=1}^m M^p_i B^x_i \sum_{i=1}^m M^x_i B^x_i\right] \\
    &= \frac{m^3 - 3m^2 + 2m}{m^3 (\eta \tau_x)^{3/2} V_t^3 \alpha^3 \cos^3 \theta} \, \mathbb{E}[p_m x_b] \, \mathbb{E}[p_m x_b] \, \mathbb{E}[x_m x_b] + \frac{2(m^2 - m)}{m^3 (\eta \tau_x)^{3/2} V_t^3 \alpha^3 \cos^3 \theta} \, \mathbb{E}[p_m x_b] \, \mathbb{E}[p_m x_b x_m x_b] \\
    &\quad + \frac{m^2 - m}{m^3 (\eta \tau_x)^{3/2} V_t^3 \alpha^3 \cos^3 \theta} \, \mathbb{E}[x_m x_b] \, \mathbb{E}[p_m x_b p_m x_b] + \frac{m}{m^3 (\eta \tau_x)^{3/2} V_t^3 \alpha^3 \cos^3 \theta} \, \mathbb{E}[p_m x_b p_m x_b x_m x_b]
\end{split}
\end{equation}

\begin{equation}
\mathbb{E}[\mathbf{x}]\mathbb{E}[\mathbf{x}\mathbf{y}]=\frac{ m^3-  m^2}{ m^3 (\eta \tau_x)^{3/2}V^3_t \alpha^3 Cos^3\theta }\mathbb{E}[p_m x_b] \mathbb{E}[p_m x_b]  \mathbb{E}[x_m x_b]+\frac{m^2}{m^3 (\eta \tau_x)^{3/2}V^3_t \alpha^3 Cos^3\theta }\mathbb{E}[p_m x_b] \mathbb{E}[p_m x_b x_m x_b]
\end{equation}
\begin{equation}
\begin{split}
    \text{Var}[\mathbf{x}\mathbf{y}] &= \frac{1}{m^4 (\eta \tau_x)^2 V_t^4 \alpha^4 \cos^4 \theta} \, \text{Var}\left[\sum_{i=1}^m M^p_i B^x_i M^x_i B^x_i\right] + \frac{1}{m^4 (\eta \tau_x)^2 V_t^4 \alpha^4 \cos^4 \theta} \, \text{Var}\left[\sum_{i=1}^m \sum_{j=1}^m M^p_i B^x_i M^x_j B^x_j (1 - \delta_{ij})\right] \\
    &= \frac{1}{m^4 (\eta \tau_x)^2 V_t^4 \alpha^4 \cos^4 \theta} \, m \, \text{Var}(p_m x_b x_m x_b) + \frac{1}{m^4 (\eta \tau_x)^2 V_t^4 \alpha^4 \cos^4 \theta} \, m (m - 1) \left[\text{Var}(p_m x_b) \, \text{Var}(x_m x_b) \right. \\
    &\quad \left. + (\mathbb{E}[x_m x_b])^2 \, \text{Var}(p_m x_b) + (\mathbb{E}[p_m x_b])^2 \, \text{Var}(x_m x_b)\right]
\end{split}
\end{equation}

% \begin{align}
%      Var[\widehat{\mathbf{T}}^{\theta}] \approx\frac{4}{m \eta \tau_x V^2_t \alpha^2 Cos^2\theta } Var[p_m x_b]+\frac{1}{m^3 (\eta \tau_x)^2 V^4_t \alpha^4 Cos^4\theta } Var[p_m x_b x_m x_b ] +\frac{m-1}{m^3 (\eta \tau_x)^2 V^4_t \alpha^4 Cos^4\theta }Var[p_m x_b]Var[x_m x_b]
% \end{align}
\begin{equation}
\begin{split}
    \text{Var}[\widehat{\mathbf{T}}^{\theta}] &\approx \frac{4}{m \eta \tau_x V_t^2 \alpha^2 \cos^2 \theta} \, \text{Var}(p_m x_b) + \frac{1}{m^3 (\eta \tau_x)^2 V_t^4 \alpha^4 \cos^4 \theta} \, \text{Var}(p_m x_b x_m x_b) \\
    &\quad + \frac{m - 1}{m^3 (\eta \tau_x)^2 V_t^4 \alpha^4 \cos^4 \theta} \, \text{Var}(p_m x_b) \, \text{Var}(x_m x_b) + \frac{m (m - 1)}{m^4 (\eta \tau_x)^2 V_t^4 \alpha^4 \cos^4 \theta} \left[(\mathbb{E}[x_m x_b])^2 \, \text{Var}(p_m x_b) + (\mathbb{E}[p_m x_b])^2 \, \text{Var}(x_m x_b)\right] \\
    &\quad + \frac{8 (m^2 - m)}{m^3 (\eta \tau_x)^{3/2} V_t^3 \alpha^3 \cos^3 \theta} \, \mathbb{E}[p_m x_b] \, \mathbb{E}[p_m x_b] \, \mathbb{E}[x_m x_b] - \frac{4(m^2 - 2m)}{m^3 (\eta \tau_x)^{3/2} V_t^3 \alpha^3 \cos^3 \theta} \, \mathbb{E}[p_m x_b] \, \mathbb{E}[p_m x_b x_m x_b] \\
    &\quad - \frac{4(m^2 - m)}{m^3 (\eta \tau_x)^{3/2} V_t^3 \alpha^3 \cos^3 \theta} \, \mathbb{E}[x_m x_b] \, \mathbb{E}[p_m x_b p_m x_b] - \frac{4m}{m^3 (\eta \tau_x)^{3/2} V_t^3 \alpha^3 \cos^3 \theta} \, \mathbb{E}[p_m x_b p_m x_b x_m x_b]
\end{split}
\end{equation}

Due to the constant positivity of $4 (\mathbb{E}[\mathbf{x}^2\mathbf{y}]-\mathbb{E}[\mathbf{x}]\mathbb{E}[\mathbf{x}\mathbf{y}] )$, the expression $4Var[\mathbf{x}]+Var[\mathbf{x}\mathbf{y}]-4 (\mathbb{E}[\mathbf{x}^2\mathbf{y}]-\mathbb{E}[\mathbf{x}]\mathbb{E}[\mathbf{x}\mathbf{y}] )$ will consistently yield a value that is lower than $4Var[\mathbf{x}]+Var[\mathbf{x}\mathbf{y}]$. In order to establish an upper limit for $Var[\widehat{\mathbf{T}}^{\theta}]$, it is acceptable to disregard the contribution of $4 (\mathbb{E}[\mathbf{x}^2\mathbf{y}]-\mathbb{E}[\mathbf{x}]\mathbb{E}[\mathbf{x}\mathbf{y}] )$.
\begin{equation}
    Var[\widehat{\mathbf{T}}^{\theta}]\approx\frac{4}{m \eta \tau_x V^2_t \alpha^2 Cos^2\theta } Var[p_m x_b]+O(1/m^2)
\end{equation}
Neglecting terms of order higher than $O(1/m^2)$, we have:
\begin{equation}
    Var[\widehat{\mathbf{T}}^{\theta}]\approx Var[\hat{\theta}]\approx \frac{ 4 V_m (1 +\eta \tau_x \epsilon  + \eta \tau_x V_m +\eta \tau_x \alpha^2 V_m Sin^2\theta)}{m \eta \tau_x V^2_t \alpha^2 Cos^2\theta } 
\end{equation}

\begin{equation}
    \theta_{up}=\theta+6.5\sqrt{\frac{ 4 V_m (1 +\eta \tau_x \epsilon  +  \eta \tau_x V_m +\eta \tau_x \alpha^2 V_m Sin^2\theta)}{m \eta \tau_x V^2_t \alpha^2 Cos^2\theta } }
\end{equation}
\begin{equation}
    \theta_{low}=\theta-6.5\sqrt{\frac{ 4 V_m (1 +\eta \tau_x \epsilon  +  \eta \tau_x V_m +\eta \tau_x \alpha^2 V_m Sin^2\theta)}{m \eta \tau_x V^2_t \alpha^2 Cos^2\theta } }
\end{equation}
% \begin{align*}
%         Var[\hat{\theta}]&=Var[\frac{1}{m\sqrt{t}V_m \alpha Cos\theta } \sum^m_{i=1}M^p_iB^x_i(2-\frac{1}{m\sqrt{t}V_m \alpha Cos\theta } \sum^m_{i=1}M^x_iB^x_i)]\\
%         &=Var[\frac{2}{m\sqrt{t}V_m \alpha Cos\theta } \sum^m_{i=1}M^p_iB^x_i]
%         +Var[\frac{1}{m^2 t V^2_t \alpha^2 Cos^2\theta } \sum^m_{i=1}M^p_iB^x_i\sum^m_{i=1}M^x_iB^x_i]\\
%         &=\frac{4}{m^2 t V^2_t \alpha^2 Cos^2\theta } Var[\sum^m_{i=1}M^p_iB^x_i]
%     -\frac{1}{m^4 t^2 V^4_t \alpha^4 Cos^4\theta }Var[\sum^m_{i=1}M^p_iB^x_i\sum^m_{i=1}M^x_iB^x_i]\\
%     &=\frac{4}{m t V^2_t \alpha^2 Cos^2\theta }  Var[p_m x_b]
%     -\frac{1}{m^4 t^2 V^4_t \alpha^4 Cos^4\theta }(m Var[p_m x_b x_m x_b ] +m (m-1)Var[p_m x_b]Var[x_m x_b] )\\
%     &=\frac{4 Var[p_m x_b]}{m t V^2_t \alpha^2 Cos^2\theta }  
%     -\frac{m (m-1)Var[p_m x_b]Var[x_m x_b]}{m^4 t^2 V^4_t \alpha^4 Cos^4\theta }(m Var[p_m x_b x_m x_b ] + \sum^m_{i=1}M^p_iB^x_i \sum^m_{i=1}M^x_iB^x_i )\\
%      &= \sum^m_{i=1} M^p_iB^x_i M^p_iB^x_i + \sum^m_{i=1}\sum^m_{j=1} M^p_iB^x_i M^p_jB^x_j (1-\delta_{ij}) 
% \end{align*}

% \begin{equation}
%     Var[\sum^m_{i=1} M^p_iB^x_i M^x_iB^x_i] + Var[\sum^m_{i=1}\sum^m_{j=1} M^p_iB^x_i M^x_jB^x_j (1-\delta_{ij})]= m Var[p_m x_b x_m x_b ] +m (m-1)Var[p_m x_b]Var[x_m x_b]
% \end{equation}

% \begin{equation}
%     Var[\hat{\theta}]=???
% \end{equation}
The channel transmission is estimated from the correlations between modulation and the outcome of the measurements at Bob's end. Conventionally, the estimator for the channel transmission is defined as $\hat{T}=\frac{(\widehat{C_{AB}})^2}{V^2_m}$. Where $\widehat{C_{AB}}$ is defined as:
\begin{equation}
    \widehat{C_{AB}}:=\frac{1}{m}\sum^m_{i=1}M_iB_i.
\end{equation}
$M_i$ is the modulation of the coherent state and $B_i$ is the corresponding measurement at Bob's end. The expression $\widehat{C_{AB}}^2/Var[\widehat{C_{AB}}]$ follows a non-central $\chi^2$ distribution with one degree of freedom and a non-central parameter $\lambda=\mathbb{E}[\widehat{C_{AB}}]^2/Var[\widehat{C_{AB}}]$. Therefore, the variance of the estimator is:
\begin{equation}
    Var[\hat{T}]=\frac{2 Var[\widehat{C_{AB}}](Var[\widehat{C_{AB}}]+2\mathbb{E}[\widehat{C_{AB}}]^2)}{V_m^4}=\frac{4 Var[\widehat{C_{AB}}] \mathbb{E}[\widehat{C_{AB}}]^2}{V_m^4}+O(1/m^2)
\end{equation}

Due to the presence of asymmetry, the conventional approach of defining the channel transmission estimator is no longer applicable. Consequently, we define the transmission estimator as follows:
\begin{equation}\hat{T}=\frac{(\widehat{C_{AB}})^2}{\alpha^2V^2_t(\sqrt{\tau_x} Cos\theta+\sqrt{\tau_p} Cos\phi)^2}.\end{equation} 
Where $\widehat{C_{AB}}$ is now defined as:
\begin{equation}
    \widehat{C_{AB}}:=\frac{1}{m}\sum^m_{i=1}M^x_iB^x_i-M^p_iB^p_i
\end{equation}
\begin{equation}
    \mathbb{E}[\widehat{C_{AB}}]=\alpha \sqrt{\eta} V_m(\sqrt{\tau_x} Cos\theta+\sqrt{\tau_p} Cos\phi)
\end{equation}
\begin{equation}
\begin{split}
    Var[\widehat{C_{AB}}]=\frac{1}{m} (Var[x_m x_b]+Var[p_m p_b])
    =\frac{1}{m}V_m (\eta \tau_x\alpha^2 V_m Cos^2\theta + \eta \tau_x\alpha^2 V_m + 
   1 + \eta \tau_x\epsilon )\\+\frac{1}{m}V_m (\eta \tau_p\alpha^2 V_m Cos^2\phi + \eta \tau_p\alpha^2 V_m + 
   1 + \eta \tau_p\epsilon )
    \end{split}
\end{equation} 
The variance of the estimator of the transmission is
\begin{equation}
    Var[\hat{T}]=\frac{Var\left[\widehat{C_{AB}}^2\right]}{\alpha^4V^4_t(\sqrt{\tau_x} Cos\theta+\sqrt{\tau_p} Cos\phi)^4}+(\mathbb{E}\left[\widehat{C_{AB}}^2\right])^2Var\left[\frac{1}{\alpha^2V^2_t (\sqrt{\tau_x} Cos\theta+\sqrt{\tau_p} Cos\phi)^2}\right]+O(1/m^2)
\end{equation}
Ignoring terms of the order $O(1/m^2)$, we have:
\begin{equation}
    Var[\hat{T}]\approx\frac{4\eta Var[\widehat{C_{AB}}]}{\alpha^2V^2_t(\sqrt{\tau_x} Cos\theta+\sqrt{\tau_p} Cos\phi)^2}+\eta^2 (\sqrt{\tau_x} Cos\theta+\sqrt{\tau_p} Cos\phi)^4Var\left[\frac{1}{ (\sqrt{\tau_x} Cos\theta+\sqrt{\tau_p} Cos\phi)^2}\right]
\end{equation}
The variance of the second term is obtained by integrating the following integrals numerically,
\begin{equation}
\begin{split}
    Var\left[\frac{1}{ (\sqrt{\tau_x} Cos\theta+\sqrt{\tau_p} Cos\phi)^2}\right]=\frac{1}{ 2 \pi Var[\widehat{\mathbf{T}}^{\theta}] Var[\widehat{\mathbf{T}}^{\phi}]}\int_{\theta_{low}}^{\theta_{up}}\int_{\phi_{low}}^{\phi_{up}}\frac{e^{-(\theta-\mathbb{E}[\theta])/2( Var[\widehat{\mathbf{T}}^{\theta}])}e^{-(\phi-\mathbb{E}[\phi])/2 Var[\widehat{\mathbf{T}}^{\phi}])}d\theta d\phi}{(\sqrt{\tau_x} Cos\theta+\sqrt{\tau_p} Cos\phi)^4}\\ -\left(\frac{1}{ 2 \pi Var[\widehat{\mathbf{T}}^{\theta}] Var[\widehat{\mathbf{T}}^{\phi}]}\int_{\theta_{low}}^{\theta_{up}}\int_{\phi_{low}}^{\phi_{up}}\frac{e^{-(\theta-\mathbb{E}[\theta])/2( Var[\widehat{\mathbf{T}}^{\theta}])}e^{-(\phi-\mathbb{E}[\phi])/2 Var[\widehat{\mathbf{T}}^{\phi}])}d\theta d\phi}{(\sqrt{\tau_x} Cos\theta+\sqrt{\tau_p} Cos\phi)^2}\right)^2
    \end{split}
\end{equation}

Estimator of channel noise $V_{\varepsilon}=t\varepsilon$ for symmetric channel noise can be defined as $\widehat{V_{\varepsilon}}=\frac{1}{2m}\sum^m_{i=1}(B^x_i+B^p_i-\sqrt{\eta}\alpha M^x_i(\sqrt{\tau_x}Cos\theta-\sqrt{\tau_p}Sin\phi)+\sqrt{\eta}\alpha M^p_i(\sqrt{\tau_p}Cos\phi-\sqrt{\tau_x}Sin\theta)))^2-2$.
The variance of the estimator can be approximated to $Var[\widehat{V_{\varepsilon}}]=\frac{2(t \varepsilon+1)^2}{m}:=s^2$.
We consider $\mathbf{E}(t^{low})=t-6.5\sigma$ and  $\mathbf{E}(V_\varepsilon^{up})=V_\varepsilon+6.5s$ corresponding to error probability of $10^{-10}$.\\
It has been demonstrated that conducting error correction prior to parameter estimation enables the utilization of all available measurements for both key generation and error correction \cite{leverrier2015composable}. However, achieving the true mutual information between Alice and Bob, denoted as $I^{(t)}_{AB}$, is not feasible because Alice lacks the necessary information about the imbalance required to execute the transformation for retrieving the genuine mutual information. To harness the benefits of this transformation, it is necessary to perform parameter estimation, at least for estimating the imbalance, prior to error correction. Consequently, we examine two potential finite key rates:

 \begin{equation}
    K_{\{n\}}=\frac{n}{N}\left[K^{(g)}_\infty\left(t_\text{low}, \varepsilon_\text{up},\delta_{up}\right)-\Delta(n)\right],
    \end{equation}
\begin{equation}
    K_{\{N\}}=K^{(s)}_\infty\left(t_\text{low}, \varepsilon_\text{up},\delta_{up}\right)-\Delta(N)
    \end{equation}

%\bibliography{lib}

\end{document}